\documentclass[letterpaper, conference,10pt] {IEEEtran}
\usepackage[ left=0.64in, right=0.64in, top=0.75in, bottom=0.75in]{geometry}
\usepackage{amsmath}
\usepackage{amssymb}
\usepackage{bm}
\usepackage{color}
\usepackage{graphicx}

\usepackage[linesnumbered,ruled,lined]{algorithm2e}

\usepackage[outdir=./]{epstopdf}
\usepackage{caption}
\usepackage{lipsum}
\usepackage{cite}
\usepackage{multirow}

\usepackage{fancyhdr}
\usepackage[caption=false, font=footnotesize]{subfig}
\SetKwInput{Input}{input}
\SetKwInput{Output}{output}

\IEEEoverridecommandlockouts
\begin{document}

	\title{Rate-Splitting for Multi-Antenna Non-Orthogonal Unicast and Multicast Transmission
		\\[-0.7ex]}

		\author{
			 
			\IEEEauthorblockN{Yijie Mao$^*$,  Bruno Clerckx$^\dagger$ and Victor O.K. Li$^*$ }
		\IEEEauthorblockA{$^*$The University of Hong Kong, Hong Kong, China\\
			$^\dagger$Imperial College London, United Kingdom\\
				Email: $^*$\{maoyijie, vli\}@eee.hku.hk, $^\dagger$b.clerckx@imperial.ac.uk}
			
			 \\[-7.5ex]
			\thanks{This work is partially supported by the U.K. Engineering and Physical
				Sciences Research Council (EPSRC) under grant EP/N015312/1.}	
			}

\maketitle
%\vspace{-300mm}
%\vspace{-100mm}

\thispagestyle{empty}
\pagestyle{empty}

\begin{abstract}
% we investigate the use of Rate-Splitting (RS) in multi-antenna non-orthogonal unicast and multicast transmission systems.
In a superimposed unicast and multicast transmission system, one layer of Successive Interference Cancellation (SIC) is required at each receiver to remove the multicast stream before decoding the unicast stream. In this paper, we show that a linearly-precoded Rate-Splitting (RS) strategy at the transmitter can efficiently exploit this existing SIC receiver architecture. By splitting the unicast message into common and private parts and encoding the common parts along with the multicast message into a super-common stream decoded by all users, the SIC is used for the dual purpose of separating the unicast and multicast streams as well as better managing the multi-user interference between the unicast streams. The precoders are designed with the objective of maximizing the Weighted Sum Rate (WSR) of the unicast messages subject to a Quality of Service (QoS) requirement of the multicast message and a sum power constraint. Numerical results show that RS outperforms existing Multi-User Linear-Precoding (MU--LP) and power-domain Non-Orthogonal Multiple Access (NOMA) in a wide range of user deployments (with a diversity of channel directions and channel strengths). Moreover, since one layer of SIC is required to separate the unicast and multicast streams, the performance gain of RS comes without any increase in the receiver complexity compared with MU--LP. Hence, in such non-orthogonal unicast and multicast transmissions, RS provides rate and QoS enhancements at no extra cost for the receivers.
\end{abstract}
\maketitle

%\let\thefootnote\relax\footnotetext[1]{This work is partially supported by the U.K. Engineering and Physical
%	Sciences Research Council (EPSRC) under grant EP/N015312/1.}
% Note that keywords are not normally used for peerreview papers.
%\maketitle
%\vspace{-7mm}

%\section{Introduction}
%\textit{Notations}: The boldface uppercase and lowercase letters are used to represent matrices and vectors, respectively. $\mathbb{C}$ denotes the complex space. $\mathbf{A}^H$ represents the conjugate transpose of matrix $\mathbf{A}$. $\mathbf{I}_{N}$ stands for the $N \times N$ identity matrix. 
%The closure of a set $\mathcal{S}$ is denoted by $\mathrm{cl} \;\mathcal{S}$ and its interior is denoted by $\mathring{\mathcal{S}} $.
\vspace{-2mm}
\section{Introduction}
\vspace{-1mm}
%The demands for multicast services, such as media streaming, mobile TV, have been growing exponentially recently. Meanwhile,  advanced wireless devices continue to strive for ever higher data rates of unicast services. As the two essential services in wireless communications, unicasting and multicasting are supported via time/frequency multiplexing.
%However, such strategy leads to inefficient resource utilization as orthogonal time/frequency resources are dedicately assigned to each service \cite{chen2017joint,tervo2017energy}. 

The scarcity of radio resources and the heterogeneity of wireless applications in 5G and beyond  motivate recent research on the non-orthogonal unicast  and multicast  transmission \cite{Zhao2016LDM,Liu2017LDM,chen2017joint,tervo2017energy}, which is based on  Layered Division Multiplexing (LDM) in the literature of digital television systems \cite{LDM2016}.  The unicast and multicast messages are precoded and superimposed at the transmitter and then broadcast to the receivers in the same time-frequency resources. Each receiver decodes and removes the multicast message using Successive Interference Cancellation (SIC) before decoding its intended unicast message. Joint unicast and multicast beamforming of multi-cell cooperative transmission has been studied, e.g., for  transmit power minimization \cite{Zhao2016LDM,Liu2017LDM},  Weighted Sum Rate (WSR) maximization \cite{chen2017joint} and energy efficiency maximization \cite{tervo2017energy}. However, all of the above works consider the use of Multi-User Linear Precoding (MU--LP) beamforming for the unicast messages. Once the common message is successfully decoded and subtracted from the received signal, each receiver fully decodes its intended unicast message by treating the interference as noise. However, MU--LP is sensitive to the user channel orthogonality and strengths. Another method is to apply power-domain Non-orthogonal Multiple Access (NOMA) to decode the unicast messages at the cost of more layers of SICs. Power-domain NOMA relies on Superposition Coding (SC) at the transmitter and SIC at the receivers \cite{NOMA2013YSaito}. It is denoted in short as SC--SIC in this work. By using SC--SIC, some users are forced to fully decode and cancel interference created by other users. Such method is only suitable when the user channels are (semi-) aligned and exhibit a large disparity of strengths.

In contrast to MU--LP and SC--SIC, linearly-precoded Rate-Splitting (RS) is an emerging multi-user multi-antenna transmission strategy  where  each unicast message is split into a common part and  a private part at the transmitter \cite{RSintro16bruno}. The common part is required to be decoded by all the receivers and removed from the received signal using SIC before each receiver decodes its intended private part by treating the interference from other users as noise. RS can be viewed mathematically as a non-orthogonal unicast and multicast transmission strategy given the superimposed transmission of common and private messages. Hence, RS was termed joint multicasting and broadcasting in \cite{hamdi2015multicasting}. However, common message in RS has an objective different from that of a conventional multicast message. The multicast message is intended and decoded by all the users while the common message of RS is decoded by all users but is intended to a subset of users. Its presence enables the decoding of part of the multi-user interference and treating the remaining part of the interference as noise.

In this work, motivated by the benefits of RS in multi-antenna Broadcast Channels (BC)  \cite{RSintro16bruno, RS2016hamdi,mao2017rate}, we  propose the use of RS in non-orthogonal  unicast and multicast transmissions. To the best of our knowledge, this is the first work that applies RS to non-orthogonal  unicast and multicast transmissions. In such a setup, we split the unicast messages into common and private parts and encode the common parts along with the multicast message into a super-common stream decoded by all users. A single layer of SIC in RS is then used for the dual purpose of separating the unicast and multicast streams as well as better managing the multi-user interference of the unicast streams. We design the precoders by formulating the WSR maximization problem of the unicast messages with a Quality of Service (QoS) requirement of the multicast message and a sum power constraint. The problem is transformed into an equivalent Weighted  Minimum Mean Square Error (WMMSE) problem and solved using an Alternating Optimization (AO) algorithm. We demonstrate in the numerical results that the rate region of RS is always equal to or larger than that of MU--LP and SC--SIC. Importantly, this performance gain comes at no additional cost for the receivers since one layer of SIC is required to separate unicast and multicast streams in the conventional MU--LP strategy. In other words, RS makes a better use of the existing SIC architecture.
%By encoding the common parts of the unicast messages with the multicast message into a common stream, only one layer of SIC is required at each receiver to decode the intended messages.
%We also argue that applying RS to the non-orthogonal  unicast and multicast transmission boosts the system throughput but maintains the same  receiver complexity as in MU--LP. 

\par The rest of the paper is organized as follows. In Section \ref{sec: mulp}, the existing MU--LP beamforming is overviewed. The proposed RS beamforming and the optimization framework are respectively specified in Section \ref{sec: RS} and Section \ref{sec: algorithm}. Section \ref{sec: simulation} illustrates numerical results and Section \ref{sec: conclusion} concludes the paper. 

\par \textit{Notations}: $\mathbb{C}$ and $\mathbb{E}\{\cdot\}$ respectively refer to the complex space and the statistical expectation.  The boldface uppercase and lowercase letters represent matrices and vectors, respectively. $\left\Vert\cdot\right\Vert$ is the Euclidean norm. 
The superscripts $(\cdot)^T$ and $(\cdot )^H$  correspond to transpose
and conjugate-transpose operators.
$\mathrm{tr}(\cdot)$ and $\mathrm{diag}(\cdot)$ are the trace and diagonal entries.

\section{Existing MU--LP Beamforming}
\label{sec: mulp}
\par In this work, we consider a BS equipped with $N_t$ antennas serving $K$ single-antenna users. The users are indexed by the set $\mathcal{K}=\{1,\ldots,K\}$. In each time frame, the BS wants to transmit a multicast message $W_0$ intended for all users and $K$ unicast messages $W_1,\ldots,W_K$ intended for different users.  The messages $W_0,W_1,\ldots,W_K$ are independently encoded into data streams $s_0,s_1,\ldots,s_K$.  The stream vector $\mathbf{s}=[s_0,s_1,\ldots,s_K]^{T}$ is precoded using the precoder $\mathbf{P}=[\mathbf{p}_0,\mathbf{p}_1,\ldots,\mathbf{p}_K]$, where $\mathbf{p}_0\in\mathbb{C}^{N_t\times1}$ and $\mathbf{p}_k\in\mathbb{C}^{N_t\times1}$ are the respective precoders of the multicast stream $s_0$ and the unicast stream $s_k$, $\forall k \in \mathcal{K}$. Assuming that $\mathbb{E}\{\mathbf{{s}}\mathbf{{s}}^H\}=\mathbf{I}$, the transmit power is constrained by  $\mathrm{tr}(\mathbf{P}\mathbf{P}^{H})\leq P_{t}$.   The resulting transmit signal $\mathbf{x}\in\mathbb{C}^{N_t\times1}$  is given by
\vspace{-1mm}
\begin{equation}
\mathbf{x}=\mathbf{P}\mathbf{{s}}={\mathbf{p}_{0}s_{0}}+\sum_{k\in\mathcal{K}}\mathbf{p}_{k}s_{k}.
\vspace{-2mm}
\end{equation}

The signal received at user-$k$  is
$
y_{k}=\mathbf{{h}}_{k}^{H}\mathbf{{x}}+n_{k},
$
where $\mathbf{{h}}_{k}\in\mathbb{C}^{N_{t}\times1}$ is the channel between the BS and user-$k$. $n_{k}\sim\mathcal{CN}(0,\sigma_{n,k}^{2})$ is the Additive White Gaussian Noise (AWGN) at user-$k$. Without loss of generality, we assume the noise variances are equal to one. The transmit SNR is equal to the total power consumption $P_t$. We assume perfect Channel State Information at the Transmitter (CSIT) and perfect Channel State Information at the Receivers (CSIR). 

At user sides, each user first decodes the multicast stream by treating the signal of all  the unicast streams as interference. The Signal-to-Interference-plus-Noise Ratio (SINR) of the multicast stream at user-$k$ is
$
\label{eq: common sinr}
\gamma_{k,0}={\left|\mathbf{{h}}_{k}^{H}\mathbf{{p}}_{0}\right|^{2}}/{(\sum_{j\in\mathcal{K}}\left|\mathbf{{h}}_{k}^{H}\mathbf{{p}}_{j}\right|^{2}+1)}.
$
Once  $s_{0}$ is successfully decoded, its contribution to the  original received signal  $y_k$ is subtracted. After that, user-$k$ decodes its unicast stream $s_{k}$ by treating the unicast streams of other users as noise. The SINR of decoding the unicast stream $s_{k}$ at user-$k$ is
$
\label{eq: private sinr}
\gamma_{k}={\left|\mathbf{{h}}_{k}^{H}\mathbf{{p}}_{k}\right|^{2}}/{(\sum_{j\in\mathcal{K},j\neq k}\left|\mathbf{{h}}_{k}^{H}\mathbf{{p}}_{j}\right|^{2}+1)}.
$
The corresponding achievable rates of $s_{0}$ and $s_{k}$ at user-$k$ are
\vspace{-2mm}
\begin{equation}
\label{eq: individual rate}
R_{k,0}=\log_{2}\left(1+\gamma_{k,0}\right)\, \textrm{and}\,\,
R_{k}=\log_{2}\left(1+\gamma_{k}\right).
\vspace{-2mm}
\end{equation}
To ensure that $s_{0}$ is successfully decoded by all users, the achievable  rate of $s_0$ shall not exceed 
\vspace{-1mm}
\begin{equation}
\label{eq: common rate}
R_{0}=\min\left\{ R_{1,0},\ldots,R_{K,0}\right\}.
\vspace{-2mm}
\end{equation} 

In this work, we maximize the WSR of the unicast messages while the rate constraint of the multicast message and the power constraint of the BS should be met. For a given weight vector $\mathbf{u}=[u_1,\ldots,u_K]$, the WSR achieved by the unicast messages in the $K$-user MU--LP assisted multi-antenna non-orthogonal unicast and multicast is
	\vspace{-2mm}
\begin{subequations}
	\label{eq: mulp}
	\begin{align}
	R_{\mathrm{MU-LP}}(\mathbf{u})&=\max_{\mathbf{{P}}}\sum_{k\in\mathcal{K}}u_{k}R_{k}  \\
	\mbox{s.t.}\quad
	& R_{k,0}\geq R_0^{th},\forall k\in\mathcal{K}\\
	&	\text{tr}(\mathbf{P}\mathbf{P}^{H})\leq P_{t}
		\vspace{-2mm}
	\end{align}
	
\end{subequations}
where   $R_0^{th}$ is the lower bound on the multicast rate.

\section{Proposed Rate-splitting Beamforming}
%\section{System Model and Problem Formulation}
\label{sec: RS}
 In this section, we introduce the use of RS to the system. We highlight the difference between the proposed RS beamforming and the existing MU--LP beamforming. The contents that are not specified remain consistent with Section \ref{sec: mulp}. 
 
% The $K$-user RS assisted multi-antenna non-orthogonal unicast and multicast transmission model  
 \begin{figure}[t!]
 	\centering
 	\includegraphics[width=3.0in]{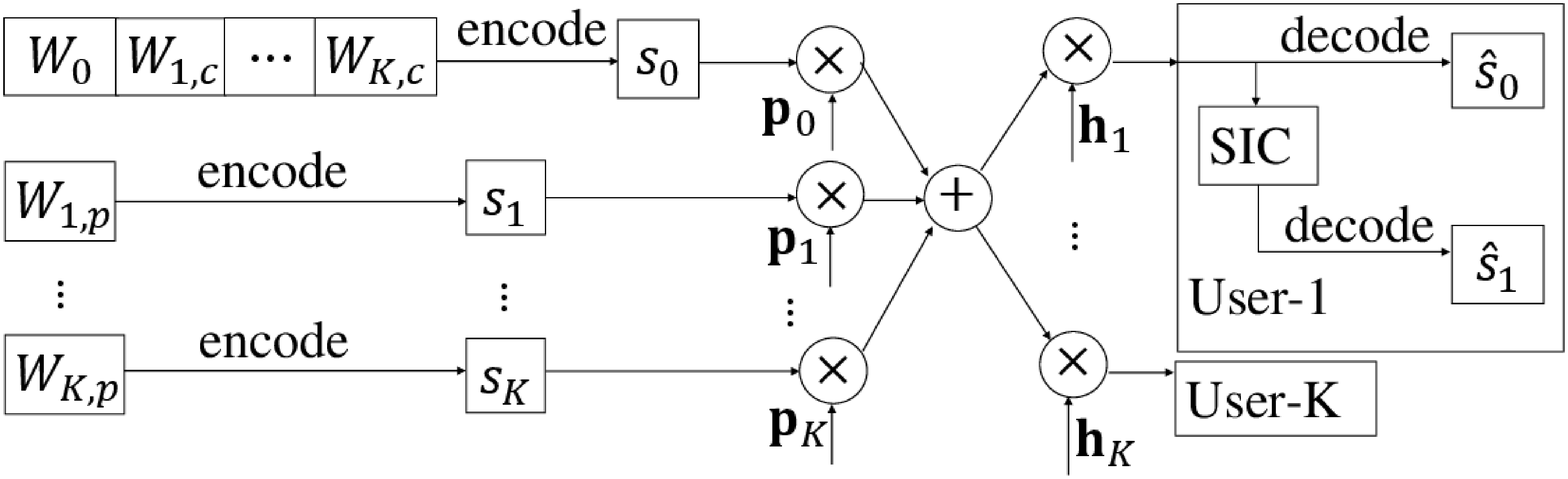}%
 	\caption{$K$-user RS assisted multi-antenna non-orthogonal unicast and multicast transmission model}
 	\label{fig: transmission model}
 	 \vspace{-5.5mm}
 \end{figure}

 The main difference between MU--LP  and RS beamforming lies in the generation of the data streams. 
  Different from MU--LP beamforming where the messages are directly encoded into independent streams, the unicast message $W_k$ intended for user-$k$ is split into a common part $W_{k,c}$ and a private part $W_{k,p}$, $\forall k\in\mathcal{K}$. The common parts of the unicast messages $W_{1,c},\ldots,W_{K,c}$ are encoded along with the multicast message $W_0$ into a super-common stream ${s}_0$. It is required to be decoded by all users. Note that ${s}_0$ includes not only the whole multicast message, but parts of the unicast messages intended for different users.  The private parts of the unicast streams $W_{1,p},\ldots,W_{K,p}$ are independently encoded into the private streams  ${s}_1, \ldots,{s}_K$. The stream vector $\mathbf{s}=[s_0,s_1,\ldots,s_K]^T$ are linear precoded via the precoder matrix $\mathbf{P}=[\mathbf{p}_0, \mathbf{p}_1,\ldots,\mathbf{p}_K ]$ and broadcast to users. The system model is illustrated in Fig. \ref{fig: transmission model}.  

At user sides, the super-common stream and private streams are decoded using SIC as in MU--LP. The achievable rate of the super-common stream  $R_{k,0}$  and the private stream $R_k,\forall k\in\mathcal{K}$ are calculated based on equation (\ref{eq: individual rate}).  To ensure that the super-common stream $s_{0}$ is successfully decoded by all users, the achievable  super-common rate  $R_{0}$  is calculated by equation (\ref{eq: common rate}). $R_{0}$ is shared by the rate of transmitting the multicast message $W_0$ and the  rates of transmitting the common parts of the unicast messages of all users, $W_{1,c},\ldots,W_{K,c}$. Denote $C_0$  as the portion of $R_{0}$ transmitting $W_{0}$ and  $C_{k,0}$ as the user-$k$'s portion of $R_{0}$ transmitting $W_{k,c}$, the achievable super-common rate is equal to
\begin{equation}
	C_{0}+\sum_{k\in\mathcal{K}}C_{k,0}=R_{0} 
\end{equation}
Following the RS structure described above, 
the total achievable rate of the unicast message of user-$k$ is 
$
R_{k,tot}=C_{k,0}+R_{k}.
$
For a given weight vector $\mathbf{u}$, the WSR achieved by the private messages in the $K$-user RS assisted multi-antenna non-orthogonal unicast and multicast is
\begin{subequations}
	\vspace{-1mm}
	\label{eq: rs}
	\begin{align}
	R_{\mathrm{RS}}(\mathbf{u})&=\max_{\mathbf{{P}}, \mathbf{c}}\sum_{k\in\mathcal{K}}u_{k}R_{k,tot}  \label{o1}\\
	\mbox{s.t.}\quad
	& C_{0}+\sum_{k\in\mathcal{K}}C_{k,0}\leq R_{k,0},\forall k\in\mathcal{K}\\
	& C_{0}\geq R_0^{th}\\
	& C_{k,0}\geq 0,\forall k\in\mathcal{K}\\
	&	\text{tr}(\mathbf{P}\mathbf{P}^{H})\leq P_{t} \label{constraint3}
	\end{align}
\end{subequations}
where  $\mathbf{c}=[C_0,C_{1,0},\ldots,C_{K,0}]$ is the common rate vector.  %$R_{k,0}$ is the achievable rate of the super-common message at user-$k$. 
We note that problem (\ref{eq: rs}) boils down to problem (\ref{eq: mulp}) when no power is allocated to the common messages $W_{1,c},\ldots,W_{K,c}$. Hence, RS always achieves the same or superior performance to MU--LP.

In the unicast-only transmission,  RS requires one layer of SIC to decode the common parts of the unicast messages for the purpose of enabling the capability of decoding part of the multi-user interference and treating part of the multi-user interference as noise \cite{RSintro16bruno, RS2016hamdi,mao2017rate}.  In comparison, MU--LP does not require any SIC at each receiver in the unicast-only transmission. However, one-layer of SIC is necessary at each user to decode  the multicast stream before decoding the intended unicast stream in the MU--LP assisted joint unicast and multicast transmission and RS still requires one layer of SIC at no extra cost for the receivers. The SIC of RS in the joint unicast and multicast transmission is used for separating the unicast and multicast streams as well as better managing the multi-user interference between the unicast streams.
\vspace{-1mm}
\section{Optimization Framework}
%\vspace{-1mm}
\label{sec: algorithm}
\par  The WMMSE algorithm to solve the sum rate maximization problem in RS without a multicast message is proposed in \cite{RS2016hamdi}. It is extended to solve the problem (\ref{eq: mulp}) and (\ref{eq: rs}). We firstly explain the procedure to solve the RS problem (\ref{eq: rs}) and then specify how (\ref{eq: mulp}) can be solved correspondingly.

%\subsection{WMMSE algorithm}
\par User-$k$ decodes  $s_{0}$ and $s_{k}$ sequentially using SICs. The common stream $s_{0}$ is decoded first. By using the  equalizer $g_{k,0}$,  $s_{0}$  is estimated as $\hat{s}_{0}=g_{k,0}y_k$. $ s_{k}$ is estimated using the equalizer $g_{k}$ as $\hat{s}_{k}=g_{k}(y_k-\mathbf{h}_k^H\mathbf{{p}}_{0}s_{0})$ after $s_{0}$ is successfully decoded and removed from $y_k$.  The Mean Square Errors (MSEs) of the common and private streams are defined as $\varepsilon_{k,0}\triangleq\mathbb{E}\{|\hat{s}_{k,0}-s_{k,0}|^{2}\}$ and  $\varepsilon_{k}\triangleq\mathbb{E}\{|\hat{s}_{k}-s_{k}|^{2}\}$, respectively. They are given by
\begin{equation}
	\vspace{-1mm}
\label{eq:MSE}
		\begin{aligned}
		&\varepsilon_{k,0}=|g_{k,0}|^2T_{k,0}-2\Re\{g_{k,0}\mathbf{h}_k^H\mathbf{p}_{0}\}+1,\\
		&\varepsilon_{k}=|g_{k}|^2T_k-2\Re\{g_{k}\mathbf{h}_k^H\mathbf{p}_k\}+1,
		\end{aligned}
	\end{equation}
	where $T_{k,0}\triangleq|\mathbf{h}_k^H\mathbf{p}_{0}|^2+\sum_{j\in\mathcal{K}}|\mathbf{h}_k^H\mathbf{p}_{j}|^2+1$  and  $T_{k}\triangleq T_{k,0}-|\mathbf{h}_k^H\mathbf{p}_{0}|^2$. 
	The optimum MMSE equalizers are then calculated by solving $\frac{\partial\varepsilon_{k,0}}{\partial g_{k,0}}=0$ and $\frac{\partial\varepsilon_{k}}{\partial g_{k}}=0$, which are given by
	\begin{equation}
	\label{eq:MMSE}
			\begin{aligned}
			&g_{k,0}^{\mathrm{MMSE}}=\mathbf{p}_{0}^H\mathbf{h}_k{T}_{k,0}^{-1},\,\,
			g_{k}^{\mathrm{MMSE}}=\mathbf{p}_k^H\mathbf{h}_k{T}_{k}^{-1}.
			\end{aligned}
		\end{equation}
		 Substituting (\ref{eq:MMSE}) into (\ref{eq:MSE}), the MMSEs are
		\begin{equation}
		\label{eq:opt MMSE}
				\varepsilon_{k,0}^{\textrm{MMSE}}\triangleq\min_{g_{k,0}} \varepsilon_{k,0} ={T}_{k,0}^{-1}{I}_{k,0},
				\varepsilon_{k}^{\textrm{MMSE}}\triangleq\min_{g_{k}} \varepsilon_{k} ={T}_{k}^{-1}{I}_{k},
			\end{equation}
			where ${I}_{k,0}=T_{k}$ and
			${I}_{k}=T_{k}-|\mathbf{h}_k^H\mathbf{p}_{k}|^2$. The SINRs of decoding $s_0$ and $s_k$ at user-$k$ can be  respectively transformed to $\gamma_{k,0}={1}/{\varepsilon_{k,0}^{\textrm{MMSE}}}-1$ and
			$\gamma_{k}={1}/{\varepsilon_{k}^{\textrm{MMSE}}}-1$ based on (\ref{eq:opt MMSE}).
			The rates become $R_{k,0}=-\log_{2}(\varepsilon_{k,0}^{\textrm{MMSE}})$ and $R_{k}=-\log_{2}(\varepsilon_{k}^{\textrm{MMSE}})$. 
			The WMSEs are given by
%			\vspace{-1mm}
			\begin{equation}
			\label{eq: augmented WMEs}
					\xi_{k,0}=u_{k,0}\varepsilon_{k,0}-\log_{2}(u_{k,0}),\,\,\xi_{k}=u_{k}\varepsilon_{k}-\log_{2}(u_{k}),
					\vspace{-1mm}
				\end{equation}
				where $u_{k,0}$ and $u_{k}$ are weights associated with the MSEs of user-$k$.  The optimum equalizers  $g_{k,0}^*=g_{k,0}^{\textrm{MMSE}}$ and
				$g_{k}^*=g_{k}^{\textrm{MMSE}}$ are then derived by solving $\frac{\partial\xi_{k,0}}{\partial g_{k,0}}=0$ and $\frac{\partial\xi_{k}}{\partial g_{k}}=0$. Substituting $g_{k,0}^{\textrm{MMSE}}$ and
				$g_{k}^{\textrm{MMSE}}$  into (\ref{eq: augmented WMEs}), we have
%				\vspace{-1mm}
				\begin{equation}
				\label{eq: augmented WMSs opt g}
						\begin{aligned}
						&\xi_{k,0}\left(g_{k,0}^{\textrm{MMSE}}\right)=u_{k,0}\varepsilon_{k,0}^{\textrm{MMSE}}-\log_{2}(u_{k,0}),\\
						&\xi_{k}\left(g_{k}^{\textrm{MMSE}}\right)=u_{k}\varepsilon_{k}^{\textrm{MMSE}}-\log_{2}(u_{k}).
						\end{aligned}
					\end{equation}
					 The optimum MMSE weights are obtained  by solving $\frac{\partial\xi_{k,0}\left(g_{k,0}^{\textrm{MMSE}}\right)}{\partial u_{k,0}}=0$  and $\frac{\partial\xi_{k}\left(g_{k}^{\textrm{MMSE}}\right)}{\partial u_{k}}=0$, which are given by
%					 \vspace{-2mm}
					\begin{equation}
					\label{eq: opt u}
							u_{k,0}^*=u_{k,0}^{\textrm{MMSE}}\triangleq(\varepsilon_{k,0}^{\textrm{MMSE}})^{-1},
							u_{k}^*=u_{k}^{\textrm{MMSE}}\triangleq(\varepsilon_{k}^{\textrm{MMSE}})^{-1}
						\end{equation}
					The Rate-WMMSE relationships are finally established by	substituting (\ref{eq: opt u}) into (\ref{eq: augmented WMSs opt g}). They are given by
						\vspace{-1mm} 
						\begin{equation}
						\label{eq: rate-wmmse}
								\begin{aligned}
								&\xi_{k,0}^{\textrm{MMSE}}\triangleq\min_{u_{k,0},g_{k,0}}\xi_{k,0}=1-R_{k,0},\\
								&\xi_{k}^{\textrm{MMSE}}\triangleq\min_{u_{k},g_{k}}\xi_{k}=1-R_{k}.
								\end{aligned}
										\vspace{-1mm}
							\end{equation}
							
          Based on the Rate-WMMSE relationships in (\ref{eq: rate-wmmse}),  the optimization problem (\ref{eq: rs}) is transformed equivalently  into the WMMSE problem given by
%          \vspace{-2mm} 
							\begin{subequations}
								\label{eq: rs wmmse}
								\begin{align}
							&\min_{\mathbf{{P}}, \mathbf{x},\mathbf{u},\mathbf{g}} \sum_{k\in\mathcal{K}}u_{k}\xi_{k,tot} \label{object wmmse}\\
								\mbox{s.t.}\quad
								&X_{0}+\sum_{k\in\mathcal{K}}X_{k,0}+1\geq \xi_{k,0}, \forall k\in\mathcal{K}\\
								& X_{0}\leq-R_0^{th}\\
								&  X_{k,0}\leq 0, \forall k\in\mathcal{K}\\
								&	\text{tr}(\mathbf{P}\mathbf{P}^{H})\leq P_{t} 
								\end{align}
							\end{subequations}
							where $\mathbf{x}=[X_{0},X_{1,0},\ldots,X_{K,0}]$ is the transformation of the common rate $\mathbf{c}$. $\mathbf{u}=[u_{1,0},\ldots,u_{K,0},u_{1},\ldots,u_{K}]$ and $\mathbf{g}=[g_{1,0},\ldots,g_{K,0},g_{1},\ldots,g_{K}]$ are the weights and equalizers, respectively. $\xi_{k,tot}=X_{k,0}+\xi_{k},\forall k\in\mathcal{K}$.

							\par Denote $\mathbf{u}^{\mathrm{MMSE}}$ and $\mathbf{g}^{\mathrm{MMSE}}$ as two vectors formed by the corresponding MMSE equalizers and weights. According to the KKT conditions of problem (\ref{eq: rs wmmse}), it is easy to show that $(\mathbf{u}^{\mathrm{MMSE}}, \mathbf{g}^{\mathrm{MMSE}})$ are optimal and unique.   We can obtain $(\mathbf{u}^{\mathrm{MMSE}}, \mathbf{g}^{\mathrm{MMSE}})$ 	by minimizing (\ref{object wmmse}) with respect to $\mathbf{u}$ and $\mathbf{g}$, respectively. Problem (\ref{eq: rs wmmse})  can be  transformed to problem (\ref{eq: rs}) based on the Rate-WMMSE relationship (\ref{eq: rate-wmmse}) and the common rate transformation $\mathbf{c}=-\mathbf{x}$.  The solution given by ($\mathbf{c}^*=-\mathbf{x}^*,\mathbf{P}^*$) meets the KKT optimality conditions of (\ref{eq: rs}) for any point ($\mathbf{x}^*,\mathbf{P}^*,\mathbf{u}^*,\mathbf{g}^*$) satisfying the KKT optimality conditions of (\ref{eq: rs wmmse}).  Therefore, problem (\ref{eq: rs}) and problem (\ref{eq: rs wmmse}) are equivalent.
	
%	\subsection{Alternating optimization algorithm}						
							\par  The joint optimization of ($\mathbf{x},\mathbf{P},\mathbf{u},\mathbf{g}$) in problem (\ref{eq: rs wmmse}) is non-convex.  With fixed ($\mathbf{x},\mathbf{P},\mathbf{u}$), the MMSE equalizer $\mathbf{g}^{\mathrm{MMSE}}$ is optimal. With fixed ($\mathbf{x},\mathbf{P},\mathbf{g}$), the MMSE weight $\mathbf{u}^{\mathrm{MMSE}}$ is the optimal weight. With fixed  ($\mathbf{u},\mathbf{g}$), the optimization problem (\ref{eq: rs wmmse}) is a convex Quadratically Constrained Quadratic Program (QCQP) which can be solved using interior-point methods. Hence, the AO algorithm is motivated to solve the problem.  In the $n$th iteration, the equalizers and weights are  calculated  by  $(\mathbf{u},\mathbf{g})=\left(\mathbf{u}^{\mathrm{MMSE}}(\mathbf{P}^{[n-1]}), \mathbf{g}^{\mathrm{MMSE}}(\mathbf{P}^{[n-1]})\right)$ based on the precoder $\mathbf{P}^{[n-1]}$ in the $(n-1)$th iteration.  $(\mathbf{x},\mathbf{P})$ are then calculated by solving problem (\ref{eq: rs wmmse}). $(\mathbf{u},\mathbf{g})$ and $(\mathbf{x},\mathbf{P})$ are iteratively updated until the convergence of the WSR.  Algorithm \ref{WMMSE algorithm} shows the steps of AO, where  $\epsilon$ is the error tolerance for convergence and $\mathrm{WSR}^{[n]}$ is the WSR calculated based on the updated $(\mathbf{x},\mathbf{P})$ in $n$th iteration.  Since $\mathrm{WSR}^{[n]}$ is increasing with $n$ and it is bounded above for a given power constraint, the AO algorithm is guaranteed to converge. Note that the global optimality of the solution cannot be guaranteed in general as the problem is non-convex. The initialization of the precoder $\mathbf{P}$ will influence the final results.
							
		The optimization framework described above are adopted to solve the MU--LP problem (\ref{eq: mulp}) by reformulating it into its equivalent WMMSE problem and using the AO algorithm to solve it. 
			\vspace{-3mm}
							\begin{algorithm}[h!]
								
								\textbf{Initialize}: $n\leftarrow0$, $\mathbf{P}^{[n]}$, $\mathrm{WSR}^{[n]}$\;
								\Repeat{$|\mathrm{WSR}^{[n]}-\mathrm{WSR}^{[n-1]}|\leq \epsilon$}{
									$n\leftarrow n+1$\;
									$\mathbf{P}^{[n-1]}\leftarrow \mathbf{P}$\;
									$\mathbf{u}\leftarrow\mathbf{u}^{\mathrm{MMSE}}(\mathbf{P}^{n-1})$; $\mathbf{g}\leftarrow\mathbf{g}^{\mathrm{MMSE}}(\mathbf{P}^{n-1})$\;
									update $(\mathbf{x},\mathbf{P})$ by solving (\ref{eq: rs wmmse}) using the updated $\mathbf{u}, \mathbf{g}$;

								}
								
								\caption{Alternating Optimization Algorithm}
								\label{WMMSE algorithm}

							\end{algorithm}
		\vspace{-2mm}					
%							\subsection{Imperfect CSIT scenario}
%							\par When CSIT is imperfect, the robust approach proposed in  \cite{RS2016hamdi}  is adopted to solve the problem. The precoding matrix is optimized based on the partial CSIT knowledge by solving a stochastic averaged weighted sum rate (AWSR) optimization problem. The problem is transformed into a deterministic counter part using the sample average approximation. Then we transform the AWSR problem based on the Rate-WMMSE relationship and utilize the proposed AO algorithm to find a stationary solution. The robust approach proposed in \cite{RS2016hamdi} for RS with unicast messages only  can be easily extended to solve the $K$-user RS/MU--LP assisted non-orthogonal unicast and multicast problem with imperfect CSIT based on the above optimization framework, which will not be explained here. 

%							The optimization framework can be adopted to solve the MU--LP problem by reformulating (\ref{eq: mulp}) into its equivalent WMMSE problem as
%							\begin{subequations}
%								\label{eq: mulp wmmse}
%								\begin{align}
%								R_{\mathrm{MU-LP}}(\mathbf{u})&=\arg\min_{\mathbf{{P}}, \mathbf{x},\mathbf{u},\mathbf{g}} \sum_{k\in\mathcal{K}}u_{k}\xi_{k} \\
%								\mbox{s.t.}\quad
%								& \xi_{k,0}\leq 1-R_0^{th}\\
%								&	\text{tr}(\mathbf{P}\mathbf{P}^{H})\leq P_{t} 
%								\end{align}
%							\end{subequations}
%							With the use of the AO algorithm, we are able to solve (\ref{eq: mulp}). 
	\vspace{-2mm}
\section{Numerical Results}
	\vspace{-1mm}
\label{sec: simulation}
\par  In this section,  the performance of the proposed RS beamforming is illustrated by comparing with the existing MU--LP  and  SC--SIC beamforming.  We focus on the two-user case since different two-user rate regions are easily compared in a two-dimensional figure. 

\par The SC--SIC assisted joint unicast and multicast transmission is briefed before we illustrate the results.  Different from RS where the multicast message is encoded with the common parts of the unicast messages,  the multicast message in SC--SIC is encoded along with the  unicast message  to be decoded first into a super-common stream $s_0$. Hence, the super-common rate $R_0$ of SC--SIC is shared by the rate of the multicast message as well as the rate of the unicast message to be decoded first. Following the system model of RS in Section \ref{sec: RS} and the difference between SC--SIC and RS, we can formulate the problem of SC--SIC and solve it by modifying the optimization framework. Note that the receiver complexity of SC--SIC increases with the number of users. More layers of SICs are required at each user to decode the interference from more users.
Moreover, SC--SIC is a particular instance of the proposed RS strategy when $K=2$ \cite{mao2017rate}.  RS should always achieve the same or superior performance to MU--LP and SC--SIC.

\par We investigate the influence of the  multicast rate constraint, channel strength disparity and channel angle between the users on the performance.  The BS is equipped with four antennas ($N_t=4$) serving two single-antenna users ($K=2$).  The simulation setting follows the underloaded two-user deployment in \cite{mao2017rate}. 
The channels of users are realized as $
\mathbf{h}_1=\left[1, 1, 1, 1\right]^H,
\mathbf{h}_2=\gamma\times\left[
1,e^{j\theta},e^{j2\theta},e^{j3\theta}\right]^H.
$
In the following results,  $\gamma=1$ and $\gamma=0.3$, which respectively represent equal  channel strength and  $5$ dB channel strength difference. For each  $\gamma$,  we consider four different $\theta$, $\theta\in \left[\frac{\pi }{9},\frac{2\pi }{9},\frac{\pi }{3},\frac{4\pi }{9}\right]$. When $\theta$ is less than $\frac{\pi }{9}$, the user channels are sufficiently aligned. When $\theta$ is larger than $\frac{4\pi }{9}$, the user channels are sufficiently orthogonal. The rate region is the set of all achievable points. Its boundary is calculated by varying the weights assigned to users.  We follow the weights  in  \cite{wmmse08}, where the weight of user-1 is fixed to $u_1=1$ for each weight of user-$2$ in $u_2 \in 10^{[-3, -1,-0.95,\cdots ,0.95,1, 3]}$.  The precoders of RS, MU--LP and SC--SIC are initialized using the same methods as discussed in \cite{mao2017rate}.  SNR is fixed to 20 dB.

%When CSIT is imperfect, the channels of user-1 and user-2 are realized as $
%\nonumber
%\widehat{\mathbf{h}}_{1}=\left[
%1,1,1,1\right]^H
%$
%and 
%$
%\nonumber
%\widehat{\mathbf{h}}_2=\gamma\times\left[
%1, e^{j\theta}, e^{j2\theta}, e^{j3\theta}\right]^H$, respectively. The precoders are initialized and designed using the estimated channels $\widehat{\mathbf{h}}_{1},\widehat{\mathbf{h}}_{2}$ and the same methods as stated in \cite{mao2017rate}. For the given channel estimate at the BS, the channel realization  is $\mathbf{h}_k=\widehat{\mathbf{h}}_{k}+\widetilde{\mathbf{h}}_{k},\forall k\in\{1,2\}$, where $\widetilde{\mathbf{h}}_{k}$ is the estimation error of user-$k$. $\widetilde{\mathbf{h}}_{k}$ has i.i.d. complex Gaussian entries drawn from $\mathcal{CN}(0,\sigma_{e,k}^2)$. The error covariance of user-1 and user-2 are $\sigma_{e,1}^2=P_t^{-0.6}$ and $\sigma_{e,2}^2=\gamma P_t^{-0.6}$, respectively.  1000 different channel error samples are generated for each user. Each point in the rate region is the average rate over the generated 1000 channels. 

\begin{figure}[t!]
		\vspace{-2mm}
	\centering
	\includegraphics[width=2.9in]{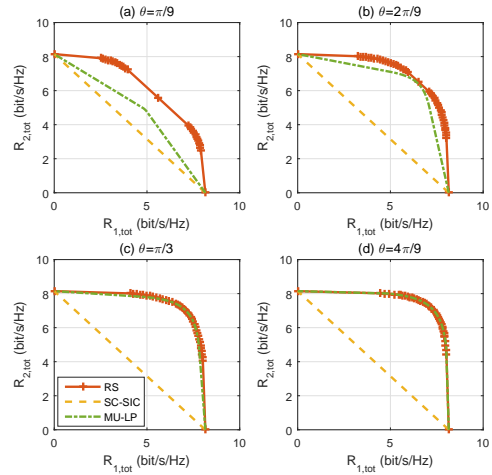}	\vspace{-3mm}
	\caption{ Achievable rate region comparison of different strategies in perfect CSIT, $\gamma=1$, $R_0^{th}=0.5$  bit/s/Hz}
	\label{fig: snr20 bias1 cth05}
	\vspace{-3mm}
\end{figure}

%\begin{figure}[t!]
%	\centering
%	\includegraphics[width=3.0in]{figure_simulation/snr20bias1cth05imperfect.eps}%
%	\caption{ Average rate region comparison of different strategies in imperfect CSIT, , $\gamma=1$, $R_0^{th}=0.5$  bit/s/Hz}
%	\label{fig: snr20 bias1 cth05 imperfect}
%\end{figure}
%
%\begin{figure}[t!]
%	\centering
%	\includegraphics[width=3.0in]{figure_simulation/snr20bias1cth15imperfect.eps}%
%	\caption{ Average rate region comparison of different strategies in imperfect CSIT, , $\gamma=1$, $R_0^{th}=1.5$  bit/s/Hz}
%	\label{fig: snr20 bias1 cth15 imperfect}
%\end{figure}
%
%\begin{figure}[t!]
%	\centering
%	\includegraphics[width=3.0in]{figure_simulation/snr20bias03cth05imperfect.eps}%
%	\caption{ Average rate region comparison of different strategies in imperfect CSIT, , $\gamma=0.3$, $R_0^{th}=0.5$  bit/s/Hz}
%	\label{fig: snr20 bias03 cth05 imperfect}
%\end{figure}

\par Fig. \ref{fig: snr20 bias1 cth05} shows the results when user-1 and user-2 have equal channel strengths ($\gamma=1$) and the multicast rate constraint is $R_0^{th}=0.5$ bit/s/Hz.  In each subfigure, the rate region achieved by RS is confirmed to be equal to or larger than that of SC--SIC and MU--LP. RS performs well for any angle between the user channels.  As the SC--SIC strategy is motivated by leveraging the channel strength difference of users, it is sensitive to the channel strength disparity. When users have equal channel strengths, SC--SIC has poor performance.  RS exhibits a clear rate region improvement over SC--SIC and MU--LP when $\theta=\frac{\pi }{9}$. The performance of MU--LP is poor when the user channels are closely aligned to each other. MU--LP is sensitive to the channel angle.
As $\theta$ increases, the gap between the rate regions of RS and MU--LP  decreases.  When $\theta=\frac{4\pi }{9}$, RS reduces to MU--LP.

\begin{figure}[t!]
	\vspace{-3mm}
	\centering
	\includegraphics[width=2.9in]{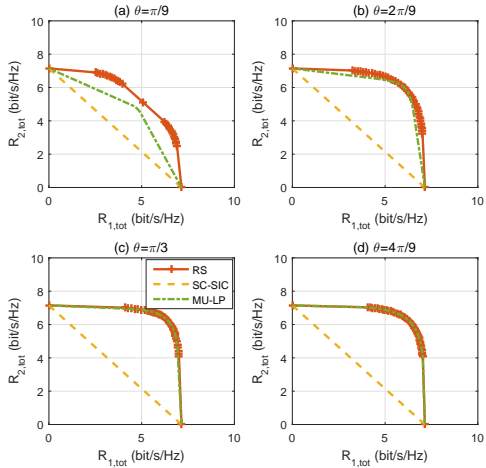}	\vspace{-3mm}
	\caption{ Achievable rate region comparison of different strategies in perfect CSIT, $\gamma=1$, $R_0^{th}=1.5$  bit/s/Hz}
	\label{fig: snr20 bias1 cth15}
	\vspace{-3mm}
\end{figure}

\begin{figure}[t!]
	\vspace{-3mm}
	\centering
	\includegraphics[width=2.9in]{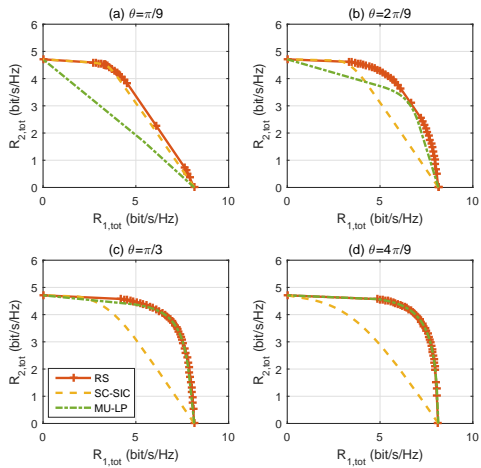}%
		\vspace{-3mm}
	\caption{ Achievable rate region comparison of different strategies in perfect CSIT, $\gamma=0.3$, $R_0^{th}=0.5$  bit/s/Hz}
	\label{fig: snr20 bias03 cth05}
	\vspace{-3mm}
\end{figure}
\par  Fig. \ref{fig: snr20 bias1 cth15} shows the results when $\gamma=1$, $R_0^{th}=1.5$ bit/s/Hz. Comparing the corresponding subfigures of Fig. \ref{fig: snr20 bias1 cth05} and Fig. \ref{fig: snr20 bias1 cth15}, the rate regions of all the strategies increase as the rate threshold of the multicast message decreases. Moreover, the rate region gain of RS over MU--LP and SC--SIC  increases as the rate threshold of the multicast message decreases. This is due to the fact the super-common stream can absorb a larger portion of the unicast messages as the rate of the multicast message decreases. RS is able to overcome the limitations of  MU--LP and SC--SIC by dynamically determining
the level of the multi-user interference to decode and treat as
noise. When the rate threshold of the multicast message decreases, more power is allocated to the unicast stream. RS exhibits further benefits of dynamic  interference management.

\par Fig. \ref{fig: snr20 bias03 cth05}  shows the results when $\gamma=0.3$, $R_0^{th}=0.5$ bit/s/Hz. Comparing with the corresponding subfigures of Fig. \ref{fig: snr20 bias1 cth05}, the rate region of SC--SIC is closer to that of RS when there is a $5$ dB channel strength difference. However, the rate region gap between RS and SC--SIC increases with $\theta$ despite the 5 dB channel strength difference. Comparing with RS, SC--SIC is more sensitive to the angle between the user channels. In Fig. \ref{fig: snr20 bias03 cth05}(b), SC--SIC and MU--LP outperform each other at one part of the rate region and the rate region of RS is larger than the convex hull of the rate regions of SC--SIC and MU--LP. We can easily draw the conclusion from Fig. \ref{fig: snr20 bias03 cth05}(b) that   RS softly bridges and outperforms MU--LP and SC--SIC. Comparing with MU--LP and SC--SIC, RS is more robust to a wide range of channel gain difference and channel angles among users. RS always outperforms MU--LP and SC--SIC. 
%Therefore, comparing with SC--SIC and MU--LP, the transmit scheduler of RS is simpler as RS is less sensitive to user deployment. Meanwhile, only one-layer of RS is required at receivers. 
This performance gain comes at no additional cost for the receivers since one layer of SIC is required to separate unicast and multicast streams in the conventional MU--LP strategy. 
\section{Conclusions}
%\vspace{-1mm}
\label{sec: conclusion}
To conclude, we exploit the benefit of the linearly-precoded  RS in the joint unicast and multicast transmission systems. Comparing with the conventional MU--LP assisted unicast and multicast transmission system where one layer of SIC is required at each receiver to remove the multicast stream before decoding the unicast stream, the proposed RS-assisted unicast and multicast transmission system further exploits the merits of the existing one layer of SIC.  By utilizing a super-common stream to encapsulate the multicast message and parts of the unicast messages, RS uses one layer of SIC to not only separate the unicast and multicast streams but also dynamically manage the multi-user interference.  We show in the numerical results that the performance of MU--LP and SC--SIC  is more sensitive to the channel strength disparity and channel angles among users. Thanks to its ability of partially decoding the interference and partially treating the interference as noise, RS softly bridges and outperforms MU--LP and SC--SIC in any user deployments.  
Moreover, the performance gain of RS increases as the rate threshold of the multicast message decreases. The benefit of RS is obtained without any increase in the receiver complexity compared with MU--LP.
Therefore, RS is a more powerful  transmission scheme for downlink multi-antenna non-orthogonal  unicast and multicast transmission systems.

\bibliographystyle{IEEEtran}
%\vspace{-3mm}
\bibliography{reference}	
%\vspace{-4mm}

\end{document}